\documentclass{mn2e}

\usepackage{graphicx}
\usepackage{graphics}
\usepackage{hhline}
\usepackage{natbib2}
\usepackage{natbibmnfix}
\usepackage{amssymb}
\usepackage{amsmath}
\usepackage{amsopn}
\usepackage{lscape}
\bibpunct[,]{(}{)}{;}{a}{}{}

\newcommand{\U}[2]{\,{\rm #1}^{#2}}

\def\adec {a_{dec(dm-\nu)}}
\def\gdec {\tilde{\Gamma}_{dec(dm-\nu)}}
\def\Tdec {T_{dec(dm-\nu)}}
\def\tdec {t_{dec(dm-\nu)}}
\def\dm {Dark Matter }
\def\dmsb {Dark Matter}
\def\cs {\sigma_{(dm-\nu)}} 
\def\bel {b_{el(dm-\nu)}}
\def\nt {\tilde{n}_{\nu}}
\def\mdm {m_{dm}}
\def\cm3 {\mbox{cm}^3 \, \mbox{s}^{-1}}

\def\beqn{\begin{eqnarray}}
\def\eqn{\end{eqnarray}}
\def\ie {\textit{i.e.} }
\def\eg {\textit{eg.} }

\def\vs {\vspace{0.2cm}}
\def\ghost#1 { }
 
\include{CelineRefs} 

\def\eg{{\it e.g.\ }}

\def\ie{{\it i.e.\ }}

\newcommand{\hkpc}{\mbox{$h^{-1}$ kpc} }

\newcommand{\hkpcKD}{\mbox{$h^{-1}$ kpc).} }
\newcommand{\hkpcDot}{\mbox{$h^{-1}$ kpc.} }
\newcommand{\hkpcCom}{\mbox{$h^{-1}$ kpc,} }
\newcommand{\hMpc}{\mbox{$h^{-1}$ Mpc} }

\newcommand{\hMpcInv}{\mbox{$h \tx{Mpc}^{-1}$} }
\newcommand{\hMpcInvDot}{\mbox{$h \tx{Mpc}^{-1}$.} }
\newcommand{\hMpcInvKet}{\mbox{$h \tx{Mpc}^{-1}$)} }

\newcommand{\hMpcDot}{\mbox{$h^{-1}$ Mpc.} }
\newcommand{\hmsun}{\mbox{$h^{-1}$ $M_{\odot}$} }
\newcommand{\hmsunKet}{\mbox{$h^{-1}$ $M_{\odot}$)} }

\newcommand{\hmsunDot}{\mbox{$h^{-1}$ $M_{\odot}$.} }

\def\la{\mathrel{\hbox{\rlap{\hbox{\lower4pt\hbox{$\sim$}}}\hbox{$<$}}}}
\def\ga{\mathrel{\hbox{\rlap{\hbox{\lower4pt\hbox{$\sim$}}}\hbox{$>$}}}}

\def\gsim{\ga}
\def\lsim{\la}

\newcommand{\bc}{\begin{center}}
\newcommand{\ec}{\end{center}}

\newcommand{\be}{\begin{equation}}
\newcommand{\ee}{\end{equation}}

\newcommand{\tx}[1] {\rmn{#1}}

\title[WIMP matter power spectra]{WIMP matter power spectra and small scale power generation} 
\author[C. Boehm, H. Mathis, J. Devriendt, J. Silk]{C. Boehm$^{1}$, 
H. Mathis$^{1}$, J. Devriendt$^{1}$ and J. Silk$^{1}$ \\
$^{1}$ DWB, Astrophysics Department, Keble Rd, OX1 3RH Oxford, UK} 

\begin{document}

\maketitle
\begin{abstract}
Dark Matter (DM) is generally assumed to be massive, cold and collisionless from the structure formation point of view. 
Although this translates into a scale-free power-law matter 
power spectrum, a more correct statement is that DM 
indeed experiences collisional damping, but on a scale which is supposed to be 
too small to be relevant for structure formation. 
The aim of this paper is to present a Cold (although ``collisional'') Dark Matter candidate whose  
matter power spectrum is damped,  discuss its consequences 
on structure formation and see whether it is distinguishable from standard candidates. 
To achieve this purpose, we first calculate the collisional damping and 
free-streaming scales of conventional (namely neutralinos) and non conventional DM candidates 
(say light particles heavier than $\sim$ 1 MeV but lighter than $O(10)$ GeV). 
The latter are supposed to annihilate to get the correct relic density and can be considered as Cold Dark Matter (CDM) 
particles in the sense that they become non relativistic before their thermal decoupling epoch. 
Unlike neutralinos, however, the linear matter power spectrum of light DM candidates can be damped on scales 
of $\sim 10^3 M_{\odot}$ (due to their interactions). 
Since these scales are of cosmological interest for structure formation, we then perform 
a series of numerical simulations to obtain the corresponding non linear matter power spectra $P(k)_{nl}$  
at the present epoch. We show that because of small scale regeneration, they all resemble each other 
at low redshifts, {\em i.e.} become very similar to a typical CDM matter power spectrum 
on all but the smallest scales. Therefore, even if lensing measurements (at redshift below unity) 
were to yield a $P(k)_{nl}$ consistent with CDM models, 
this would not constitute a sufficiently robust evidence in favour of the neutralino 
to rule out alternative DM candidates, even if their linear matter power spectrum 
is damped on relatively large scales.
\end{abstract}

\section{Introduction}

Despite decades of improvement in observations and experimental results, the nature of DM
is still under investigation. It is often claimed that the accuracy of the 
measurement of the Cosmic Microwave Background spectrum 
up to the second acoustic peak provides very strong evidence in favour of CDM scenarios. 
However other DM models, \eg Warm Dark Matter (WDM) \citep{silkshaef}, 
or scenarios in which DM has non-negligible interactions with photons (IDM) 
\citep{brhs}, also predict an identical CMB spectrum to that measured so far. 

\vs
In contrast with CDM, however, WDM particles do have a cut-off in the linear 
matter power spectrum ($P(k)_l$), at a scale fixed by the warmon mass (the 
linear matter power spectrum of IDM also has a 
cut-off but it is much less drastic than in WDM and fixed by the ratio of the elastic scattering cross section 
of DM with photons or neutrinos to its mass). 
Because of this cut-off, WDM particles lighter than $\sim$ 1 keV 
\citep{sperg} or 10 keV \citep{reio2}, which experience damping on scales smaller than  $\sim 10^9 M_{\odot}$ or 
$\thickapprox 10^4 M_{\odot}$ respectively,   
would be excluded by the WMAP detection of reionization at large redshift \citep{wmapreio}  
should it be confirmed (the major caveat being the efficiency of reionization sources see \eg \cite{julien}). 
On the other hand, it might be difficult to discriminate among typical CDM, WDM or any kind of particles 
whose linear matter power spectrum is damped at cosmological scales by using 
non-linear matter power spectra only, because of the small-scale regeneration 
mechanism  which tends to make the WDM and CDM $P(k)_{nl}$ very similar, at least 
at redshifts $z \leq 1$. Thus, the determination of the matter power spectrum by 
lensing measurements \citep{bernardeau} would probably fail to constrain the nature of DM, although 
there might exist other properties (like the structural parameters of DM haloes) 
on which lensing observations may set tighter constraints.  

\vs
Although a warmon with mass just above $\sim 10$ keV is still a viable possibility, 
CDM remains the most popular scenario, mostly because:  
\begin{itemize}
\item it is generally considered as the most simple and natural solution. Indeed, it 
does not rely on any particle physics parameter, in contrast with the 
WDM scenario where results depend on the warmon mass,
\item the most popular CDM candidate -- namely neutralinos \citep{pierre1} -- is still allowed 
by cosmology and particle physics while WDM candidates proposed so far,  
\eg gravitinos  or sterile neutrinos, appear much more constrained 
and very close to be excluded unless one invokes peculiar mechanisms \citep{murya,steen}. 
\end{itemize}
Nevertheless, it is worth to keep in mind that neutralinos (or any other well-known 
CDM particle candidate) have not been discovered 
so far, so there is room for other possibilities. 

\vs
Coming back to structure formation, DM particles are assumed 
to be fairly massive in order to avoid large free-streaming effects that would wash out the DM primordial 
fluctuations \citep{hot}. Furthermore, to prevent the Silk damping mechanism 
\citep{joenature,joe68} from operating, 
one requires that they should not have any electromagnetic interactions. This has actually led some authors 
\citep{lee,peebles} to propose, in the early eighties,  
that DM could be made of Weakly Interacting Massive particles (WIMPs), a statement 
which is now commonly accepted.  Because of this ``weakly interacting'' property, one 
generally neglects DM interactions in numerical simulations. As a result \dmsb, and WIMPs in particular,  
are said to be collisionless regarding structure formation, their 
linear matter power spectra being generally taken as a ``true'' 
scale-invariant spectrum with no cut-off.

\vs 
Obviously, such a statement is a bit too drastic. 
If they are produced thermally, 
the DM particles must have non-negligible interactions (weak or not, provided they are not electromagnetic)  
which might have consequences on structure formation.
Those interactions  turn out to be also very useful 
as they are at the basis of direct and indirect experimental detection of DM candidates \citep{direct,indirect}.  
So it is worth to try to assess their detailed impact on structure formation, instead of neglecting 
them {\em a priori}.

\vs
DM elastic scattering cross sections turn out to be important for structure formation 
because they fix the epoch at which \dm thermally decouples, \ie they specify the epoch at which 
the collisional damping stops and the free-streaming starts. 
It is therefore of crucial importance to determine the strongest 
interactions a WIMP can have and what are their effects on structure formation. 
This eventually provides a definition of the particles really permitted by structure formation 
models, thereby extending the notion of WIMPs to all particles having interactions that are not 
necessarily weak but 
small enough so as to preserve fluctuations on scales where we know they exist \citep{bfs}. 
As a result, this opens new possibilities for Cold DM candidates, bearing in mind 
that their $P(k)_l$ would then differ from a ``true'' scale invariant spectrum.  

\vs
The aim of this paper is to answer the question as to whether or not structure 
formation can distinguish standard CDM candidates like neutralinos from other more exotic 
alternatives.  To this purpose, we first review the different damping mechanisms that can 
affect DM linear matter power spectra, with  emphasis on the ``mixed'' 
damping mechanism introduced by \citet{bfs}. We then compute the damping scales of neutralino 
primordial fluctuations and compare them with those of lighter DM particles (having a mass 
$O(10\U{GeV}{}) > \mdm > O(\U{GeV}{})$). The latter indeed turn out to be viable \dm candidates 
provided their dominant annihilation cross section are S-wave suppressed so as  
to satisfy both the relic density criteria and the gamma ray/radio constraints. This condition  
can be achieved for example by introducing a ``new'' kind of interaction (not based on the exchange of 
Standard Model gauge bosons) that would have eluded experimental searches up to now \citep{bf}. 

\vs
The light DM $P(k)_l$ being potentially damped on scales relevant to structure formation, 
we perform four WDM simulations \footnote{Light DM $P(k)_l$ are in fact expected to be oscillating. 
Also one expects some power at very 
small scales (albeit much less than in CDM scenarios). Therefore, they may appear intermediate 
between standard CDM and WDM linear matter power spectra. However, we use a WDM $P(k)_l$ to perform our simulations  
so as to make the point regarding the small-scales regeneration mechanism.} 
with different scale cut-offs and one CDM numerical simulation for the sake of comparison.
We present the evolution of the non-linear matter power spectra with redshift for all these 
models in section 4 where we perform a detailed comparison of our WDM models with the typical 
standard CDM spectrum.

\vs
In light of these numerical results, we conclude that even if lensing observations favoured 
a standard CDM $P(k)_{nl}$, they would not exclude the existence of 
a cut-off in the linear matter power spectrum, unless they could probe the distribution of matter 
up to very small scales with great accuracy. Therefore, we emphasize that it is really difficult to rule out 
``exotic'' DM models on the basis of the non linear matter power spectra alone.

\section{Damping lengths}

The objects we know (\textit{e.g.} galaxies, clusters of galaxies) are thought to 
form by gravitational collapse of primordial density fluctuations. However, in order to grow significantly, these 
fluctuations need to survive well-known damping mechanisms that would wash them out. If the dominant 
species that constitutes these fluctuations was a light neutrino or any relativistic species, then the relevant damping 
mechanism would be free-streaming. This free-streaming would generate a cut-off in the linear matter power spectrum at 
very large scales and eventually turn out to be in contradiction with small-scale observations since all galaxies would 
have to be created out of pancake and/or filament fragmentation. 
A cut-off at large scale is also expected if the main component of these matter density fluctuations is baryonic matter, 
because of their very large interactions with photons; this is known as Silk damping. Since both these models 
appear in contradiction with observations, there is a need for another kind of matter, called DM, 
that would be massive enough to avoid significant free-streaming and without any ``standard'' electromagnetic 
interactions.  

\vs
In this paper, we argue that --- quite similarly to WDM scenarios where the free-streaming length 
can be as large as $10^6-10^9 \, M_{\odot}$ ---   
there exist realistic scenarios where the collisional damping lengths of CDM can be significant (at least 
up to $10^3 \, M_{\odot}$). 
These DM particles would be Cold in the sense that they annihilate and decouple after they become non-relativistic, 
but on the other hand they would have a linear matter power spectrum very different from what is 
generally expected. We also show that another damping effect, called mixed damping, can become 
particularly important in the case of light DM particles and is responsible for this peculiar behaviour.

\vs
The latter can be understood as an intermediate effect 
between collisional damping and free-streaming. 
It takes place when DM is thermally coupled to a species $i$ that is already free-streaming. 
This means that it basically starts when species $i$ thermally decouples (time 
$t_{dec(i)}$) and lasts until DM stops being in thermal equilibrium with $i$ (which 
occurs at time $t_{dec(dm-i)}$). The condition for the mixed damping to 
take place is therefore written as: $t_{dec(dm-i)} > t_{dec(i)}$. 

\vs 
In the following, we shall focus on $i=\nu$, since DM--neutrino interactions are likely to satisfy  
the previous relationship, \ie 
$$\tdec > t_{dec(\nu)}$$ 
or, equivalently, 
$$ \Tdec < T_{dec(\nu)},$$ 
where $\Tdec$ is the temperature at which the dm-$\nu$ interactions stop having an effect on 
the DM fluid  and $T_{dec(\nu)}$ is the neutrino temperature at their decoupling  
(say in a standard scenario: 
$T_{dec(\nu)} = T_{dec(\nu-e)} \sim 1 \U{MeV}{}$). 

\vs
The mixed damping effect has never been considered before because its existence has 
only been recently pointed out by \cite{bfs} (and its physical relevance by \cite{bf})  
but also because DM candidates are generally thought to 
have thermally decoupled before or at the onset of 1 MeV (\textit{i.e.} $T_{dm-\nu} > T_{dec(\nu-e)}$) 
so that the condition for this damping to occur was thought to never be satisfied.

\vs
Let us now suppose that we are in a scenario where  $\Tdec < T_{dec(\nu)}$.
The damping scale of DM primordial fluctuations then originates from    
i) the collisional damping 
effect due to the coupling of DM particles with species that are still collisional 
(this actually includes the neutrino contribution for  $T \gtrsim T_{dec(\nu)}$), 
ii) the mixed damping effect, originating from interaction with neutrinos and 
acquired during the period [ $t_{dec(\nu)}$, $\tdec$], and   finally 
iii) the DM free-streaming for DM temperatures below   $T_{dec(dm)}$ (in fact $\Tdec$ 
if $dm-\nu$ are the strongest DM interactions). 
The associated damping scales   
will be denoted $l_{cd}$, $l_{md}$, $l_{fs}$ respectively. Their expressions are given 
(according to \cite{bfs,bfs2}) by\footnote{The normalization factors are not shown in these formulae 
but we do take them into account in our calculations.}: 

\beqn
l_{cd}^2 &\propto& \int^{t_{dec(dm)}} \frac{\rho_{dm} \, v_{dm}^2}{\rho \, \Gamma_{dm} \, a^2} \, dt \, + 
\int^{t_X} \frac{\rho_{\nu} \, c^2}{\rho \, \Gamma_{\nu} \, a^2} \, dt 
\nonumber  \\ 
&&\hspace{1.cm} + \sum_{i \supset e \, , \gamma \, ..} \! \int^{t_{dec(dm-i)}} \frac{\rho_{i} \, v_{i}^2}{\rho \, \Gamma_{i} \, a^2} \, dt
\label{sv} \\ 
l_{md}^2 &\propto& \int_{t_{dec(\nu)}}^{t_{dec(dm-\nu)}} \frac{\rho_{\nu} \, c^2}{\rho \, H \, a^2} \ 
dt \, \sim \, {\left(\frac{c \, t}{a}\right)^2}_{ \vert_{\tdec}} \label{md} \\
l_{fs} &\propto& \int_{t_{dec(dm)}}^{t_0}  \frac{v}{a} \ dt \sim Max {\left(\frac{c \, t}{a}\right)}_{ \vert_{t_{dec(dm)}-t_0}}  \label{fs}
\eqn
with $i$ denoting all the species that can transmit  
their collisional damping to DM fluctuations, $\Gamma_{dm}$ the total DM interaction rate, 
$\Gamma_i = \sum_j \Gamma_{i-j}$ the sum of the partial interaction rates of species $i$ 
($j$ being the species in thermal equilibrium with $i$), $H$ the Hubble constant and 
$a = a(t)$ the scale-factor at a time $t$. Here $t_X$ denotes either $t_{dec(dm-\nu)}$
 if $t_{dec(dm-\nu)} < t_{dec(\nu)}$ (in which case there is no mixed damping since  
DM thermally decouples from neutrinos before they enter  
their free-streaming regime) or $t_{dec(\nu)}$ if  $t_{dec(dm-\nu)} > t_{dec(\nu)}$ (since in this case 
the integration of $1/\Gamma_{\nu}$ for $t > t_{dec(\nu)}$ does not have any meaning). 
We emphasize that eq.(\ref{sv}) and eq.(\ref{md}) supposes implicitly that $t_{dec(dm-i)} < t_{dec(i)}$ for $i \neq \nu$, or 
in other words, that the mixed damping species is only due to neutrinos.

\vs
We also draw attention to the fact that we wrote $l_{md}$ and $l_{cd}$ as two separate 
contributions when in fact, the mixed damping (eq.\ref{md}) is part of the collisional damping 
(eq.\ref{sv}). Thus, even if $\tdec > t_{dec(\nu)}$, we should write 
only one contribution, namely: 
$$ l_{cd(dm-\nu)}^2  \propto \int^{t_{dec(\nu)}} \frac{\rho_{\nu} \, c^2}{\rho \, \Gamma_{\nu} \, a^2} \ dt + 
\int_{t_{dec(\nu)}}^{t_{dec(dm-\nu)}} \frac{\rho_{\nu} \, c^2}{\rho \, H \, a^2} \ 
dt \ $$
with $$l_{cd}^2 \lesssim  \int^{t_{dec(dm)}} \frac{\rho_{dm} \, v_{dm}^2}{\rho \, \Gamma_{dm} \, a^2} \, dt \, 
+ \sum_{x \ \supset \ \nu \, , \gamma \, ...} l_{cd(dm-x)}^2,$$
\noindent 
but we will avoid this notation in what follows (and come back to eq.\ref{sv} and eq.\ref{md}) 
 to underline the importance of the mixed damping regime.

\vs
The reason why neutrinos are expected to be the only species of interest for the collisional and 
mixed damping can be easily understood. 
Since the transport coefficients that enter eq.\ref{sv} are proportional to 
$\frac{\rho_i \, v_i^2}{\Gamma_i}$, only photons and  neutrinos are expected to give a 
large contribution to $l_{cd}$ (and therefore to $l_{md}$). 
Their velocity is indeed maximal and they do not 
experience any annihilations that could reduce their energy density. 
To contribute to the collisional damping, however, the DM--photon interactions should be such that  
$t_{dec(dm-\gamma)} \sim t_{rec}$ (with $t_{rec}$ the recombination epoch). Such a 
condition actually requires very large (and therefore unlikely) values of 
the $dm-\gamma$ elastic scattering cross section so neutrinos should 
eventually turn out to be the only species 
of interest as far as the estimate of $l_{cd}$ and $l_{md}$ is concerned (since $t_{dec(dm-\nu)} \sim 
t_{dec(\nu)}$ is  expected to rely on interactions of weak intensity). 

\vs
In what follows, we shall therefore make the reasonable assumption that $t_{dec(dm-\gamma)} \ll t_{rec}$, and 
 neglect the photon contribution to the collisional damping effect.

\vs
Let us now compute the terms associated with neutrinos in $l_{cd}$ and $l_{md}$.  
The interaction rate $\Gamma_{\nu} = \sum_{j} \Gamma_{\nu-j}$ can be decomposed as 
$\Gamma_{\nu} = \Gamma_{\nu-e} + \Gamma_{\nu-dm} + \sum_{j \neq e} \Gamma_{\nu-j}$. 
We shall neglect the terms $\sum_{j \neq e} \Gamma_{\nu-j}$, since the $\Gamma_{\nu-j}$ are 
seen to be always smaller than $  \Gamma_{\nu-e}$, \ghost{so that 
$\Gamma_{\nu} \sim \Gamma_{\nu-e} + \Gamma_{\nu-dm}$} and 
make the assumption $\Gamma_{\nu-dm} < \Gamma_{\nu-e}$. The neutrino decoupling is then  
given by $\Gamma_{\nu-e} \sim H$, as expected in a standard scenario.  The non conventional case  
$ \Gamma_{\nu-dm} > \Gamma_{\nu-e}$ will be discussed later for completeness.

\vs
\begin{itemize}
\item Let us first study the the case where $\tdec \leq t_{dec(\nu)} \, \equiv \, t_{dec(\nu-e)}$.

\vs
This is a standard situation without mixed damping effects so we only need to 
discuss the neutrino contribution to $l_{cd}$. 
Despite the term $\rho_{\nu} \, c^2$ which appears very promising,  the maximum collisional damping 
that standard neutrinos can really transmit  
to DM fluctuations is   $ 
l_{cd(dm-\nu)}^2 \lesssim \int^{t_{dec(\nu-e)}} \frac{\rho_{\nu} \, c^2}{\rho \, \Gamma_{\nu-e} \, a^2} \- dt \ 
\sim  [O(100 \U{pc}{})]^2 \ $, 
(the maximum damping length being reached when $\tdec \, \simeq \, t_{dec(\nu-e)}$).  
Thus, if a DM candidate decouples  
from neutrinos at $\sim 1$ MeV,  the cut-off 
in the linear matter power spectrum  due to DM-neutrino collisions 
should be around $O(0.1 M_{\odot})$,  which is a remarkable property!  

\vs 
This is actually much larger than the scale computed by 
\cite{misner} (say $10^{-4}-10^{-3} \ M_{\odot}$) relative to the 
damping of electronic primordial fluctuations\footnote{The existence of DM was actually  not 
considered at that epoch.}. But one can check that his computation, now rewritten in terms of the 
Weinberg formulation \citep{weinberg}, indeed corresponds to 
$l_{Misner} \propto \int^{t_{dec(\nu-e)}} \frac{\rho_{\nu} \, c^2}{\rho \, \Gamma_{\nu-e} \, a^2} \ dt \ $, 
so that $l_{Misner} \equiv l_{cd} \simeq 100 \U{pc}{}$.

\vs
This damping  has to be compared with the free-streaming motion acquired by the DM during the period [$t_{dec(dm)}$, $t_0$] 
($t_0$ denoting the present epoch and $t_{dec(dm)} \equiv t_{dec(dm-\nu)} \sim t_{dec(\nu-e)}$ 
if $dm-\nu$ is indeed the last DM interaction). 
For annihilating particles heavier than a few MeV, which is 
a reasonable assumption in the case of a usual WIMP candidate\footnote{The damping scale of annihilating WIMPs, 
having a mass above a few MeV, is 
$l_{fs} \simeq O(200) \, \U{kpc}{} \, \left(\frac{\mdm}{\U{MeV}{}}\right)^{-1/2} \, 
\left(\frac{a_{dec(dm)}}{10^{-4}}\right)^{1/2}$,  or 
equivalently $M_{fs} \, \simeq \, 10^9 \, M_{\odot} \, \left(\frac{\mdm}{\U{MeV}{}}\right)^{-3/2} \, 
\left(\frac{a_{dec(dm)}}{10^{-4}}\right)^{3/2} \ , \ $ 
$\ a_{dec(dm)}$ being the scale-factor when the DM thermally decouples.} decoupling at $t_{dec(\nu-e)}$,   
we find a free-streaming mass of $$M_{fs} \lesssim  \left(\frac{\mdm}{\U{MeV}{}} \frac{10^{10}}{10^{4}} \right)^{-3/2} \,  
 10^9 M_{\odot} \sim \left(\frac{\mdm}{\U{MeV}{}}\right)^{-3/2} \, M_{\odot},$$ 
so $M_{cd} \sim 0.1 \, \left(\frac{\mdm}{\U{MeV}{}}\right)^{3/2} \, M_{fs}$ for 
WIMPs decoupling as late as $t_{dec(\nu-e)}$ \cite{bfs}.

\vs
\vs
\item Let us now consider the case $\tdec > t_{dec(\nu)} \equiv t_{dec(\nu-e)}$.

\vs
Since the mixed damping becomes relevant, 
the DM primordial fluctuations are expected to be washed out by 
neutrino free-streaming motion (instead of neutrino collisions). Free-streaming being extremely efficient, one 
expects a significant damping effect for a temperature $\Tdec$ just slightly below $T_{dec(\nu-e)}$ (assuming 
standard neutrinos \ie $T_{dec(\nu)} = T_{dec(\nu-e)}$).
But can $\Tdec$ be lower than $T_{dec(\nu)} \equiv T_{dec(\nu-e)}$ for ordinary WIMPs?

\end{itemize}
 
\vs 
What determines whether the mixed damping regime exists or not  
 is basically  the strength of the $dm-\nu$ interactions. The stronger they are, the lowest $\Tdec$ is, 
and the more likely the condition $ \Tdec < T_{dec(\nu)}$ is going to be fulfilled. 
Accordingly, we now proceed to establish the relationship between $\Tdec$ and the
interaction rate $\Gamma_{dm-\nu}$. 

\subsection{Relationship between $T_{dec(dm)}$ and $\Gamma_{dm-\nu}$}
\vs
The temperature can be written as $\Tdec \simeq T_0/\adec$ where $\adec  = (\gdec/\tilde{H}_r)$ is the scale-factor at 
$\tdec$, $\gdec = \Gamma_{dec(dm-\nu)} \ \adec^3$ is the effective (comoving) interaction rate 
at the same epoch and $\tilde{H}_r \sim \, 2 \, 10^{-20} \, \U{s}{-1}$ (in the radiation dominated era, which is the 
relevant epoch for our purpose).

\vs 
The relationship between the interaction rate $\Gamma_{(dm-\nu)}$ 
and the elastic scattering cross section $\cs$ between DM and neutrinos can be obtained by writing   
the Euler equations for the DM and neutrino fluids (in the Newtonian gauge) \citep{brhs}: 
\beqn
\dot{v_{dm}} &=& k \Phi - H v_{dm} - S^{-1} \, a \ \dot{\mu} \ (v_{dm} - v_{\nu})  \nonumber\\
\dot{v_{\nu}} &=& k \Phi + \frac{1}{4} k \delta_{\nu} - \frac{1}{6} k \pi_{\nu} - 
a \ \dot \kappa \ (v_{\nu} - v_b)  \nonumber \\ 
&& \hspace{1.5cm} - a \ \dot{\mu} \ (v_{\nu} - v_{dm}), \nonumber \\
&& \hspace{-1.5cm} \mbox{with} \nonumber \\
&& \hspace{-1.5cm}  \dot{\mu} = \, \cs \, c \, n_{dm}, \ \dot{\kappa} = \, \sigma_{(\nu-e)} \, c \, n_{e}, \ 
 S^{-1} \propto \frac{\rho_{\nu}}{{\rho_{dm}}}.  \nonumber
\eqn
\noindent 
Here $H$ is the Hubble parameter, $\Phi$ the Bardeen potential, 
$\delta_{\nu}$ the neutrino density perturbations, $\pi_{\nu}$ 
the anisotropic stress tensor and $v_{dm}, \ v_{\nu}$ the velocity divergence of DM and neutrinos respectively.  
We then get:
\beqn
\gdec  &=& S^{-1} \ \dot{\mu} \ a^{3}\vert_{t_{dec(dm-\nu)}}\label{gdec} \\
&=& \cs \ c \ \nt \ \Tdec/\mdm \nonumber 
\eqn
\noindent 
which implies, if we parameterize the cross section as $$\cs c = \bel \ T^n \ \ $$ ($n \geq 0$ ) 
and use the definition $\Tdec \propto \adec^{-1}$: 

\beqn
\gdec &=& \bel  \, \nt \Tdec^{n+1}/\mdm, \nonumber \\ 
\nonumber \\ 
\Tdec &=& \left(\frac{T_0 \, \tilde{H}_r \, \mdm}{ \bel  \, \nt}\right)^{1/(n+2)} \label{Tdec} \\ 
\adec &=& \left(\frac{ \bel  \, \nt \, {T_0}^{n+1}}{\, \tilde{H}_r \, \mdm}\right)^{1/(n+2)} \label{adec} \\
\cs c &=&  \ \bel \ \Tdec^n \nonumber \\
&=& \left(\frac{\bel^{\frac{2}{n}} \, T_0 \, \tilde{H}_r \, \mdm}{ \, \nt}\right)^{\frac{n}{n+2}}\label{csv}
\eqn 
Here  $\tilde{n}_{\nu} = n_{\nu} \, a^3$ is the comoving number 
density and $T/\mdm$ a factor that reflects the efficiency of the $dm-\nu$ collisions 
in transmitting the neutrino momentum to the DM. In other words, our effective 
$\Gamma_{(dm-\nu)}$ corresponds to a standard interaction rate (usually defined as 
the product of $\sigma v$ times a number density)
times the transferred momentum and $\tilde{\Gamma}_{dec(dm-\nu)}$ is the same interaction  
rate, but comoving and estimated at $t_{dec(dm-\nu)}$.

\vs
As expected, one can check from eq.\ref{Tdec}  that 
the largest the cross section (and incidentally $\bel$) is, the lowest $\Tdec$ is. 
For $n=0$, $\adec \sim 2.3 \ 10^6 \ \sqrt{\cs c} \ \left(\frac{\mdm}{\U{MeV}{}}\right)^{-1/2}$ 
(assuming that $\cs$ indeed yields a decoupling in the radiation dominated era).

\vs
For $n=2$ (the most likely value), we find: 
\beqn
\Tdec &\sim& 2.5 \ 10^{-3} \ b_{el}^{-1/4} \  \left(\frac{\mdm}{\U{MeV}{}}\right)^{1/4} \U{cm}{-1},\label{calc1} \\
\adec &\sim& 5 \ 10^3 \  b_{el}^{1/4} \,   \left(\frac{\mdm}{\U{MeV}{}}\right)^{-1/4},\nonumber \\
\cs c &\sim& 6 \ 10^{-6} \ b_{el}^{1/2} \,  \left(\frac{\mdm}{\U{MeV}{}}\right)^{1/2} \, \U{cm}{3} \U{s}{-1} \label{calc3}
\eqn 
\noindent (with $b_{el} \equiv \left(\bel/\U{cm}{5} \U{s}{-1}\right)$) 

\vs

\subsection{Conditions for  mixed damping to exist}

\vs
By using the definition \ref{gdec}, we actually introduce a ``non standard'' 
notation where, in fact, $\tilde{\Gamma}_{dec(\nu-dm)} \neq \gdec$.  The relationship between 
these two interaction rates can be written as: 
$\tilde{\Gamma}_{dec(\nu-dm)} = \frac{\rho_{dm}}{\rho_{\nu}} \gdec$
with $\rho_{dm}$  and $\rho_{\nu}$ being the energy densities of DM and neutrinos respectively. 
Roughly, one can write $\tilde{\Gamma}_{(\nu-dm)} = \sigma_{dm-\nu} \, c \, \tilde{n}_{dm} $ and 
$\tilde{\Gamma}_{(dm-\nu)} = \sigma_{dm-\nu} \, c \, \tilde{n}_{\nu} \, \frac{T}{\mdm}$.

\vs
If one requires neutrinos to decouple at $\sim$ 1 MeV (which corresponds to the 
standard scenario) and DM to be still influenced by neutrinos, then the condition for the 
mixed damping effect to take place should in fact be rewritten (still assuming monotonic 
behaviour) as: 
\beqn
\Gamma_{\nu-dm \vert_{t_{dec(\nu-e)}}} &<& \Gamma_{\nu-e \vert_{t_{dec(\nu-e)}}}  \label{cmd1} \\
\Gamma_{\nu-e \vert_{t_{dec(\nu-e)}}} &\lesssim& \Gamma_{dm-\nu \vert_{t_{dec(\nu-e)}}},  \label{cmd2} 
\eqn
\noindent 
which means, with our notation:
\beqn
\sigma_{dm-\nu} &<& \sigma_{\nu-e} \, \left({\frac{n_e}{n_{dm}}}\right)_{ \vert_{t_{dec(\nu-e)}}}\nonumber \\
\sigma_{dm-\nu} &>& \sigma_{\nu-e} \, \left({\frac{n_e}{n_{\nu}}} \, \frac{\mdm}{T}\right)_{ \vert_{t_{dec(\nu-e)}}} 
\nonumber 
\eqn
\noindent or (if one assumes that DM has finished annihilating before (or at the onset of) $t_{dec(\nu-e)}$): 
\beqn
\sigma_{dm-\nu } &<& A \ \sigma_{\nu-e} \, x_d^{-3/2} \ e^{x_d} \label{up} \\
\sigma_{dm-\nu} &\gtrsim& \sigma_{\nu-e}  \, \left(\frac{\mdm}{\U{MeV}{}} \, \frac{g_e}{g_{\nu}}\right).  \label{down}
\eqn
with $A \sim \left(\frac{g_e}{g_{dm}}\right) \frac{\zeta(3)}{\pi^2} (2 \pi)^{3/2}$ 
($\zeta(3)$ the Riemann zeta function of 3) and $x_d = m_{dm}/T_{fo} \in [12-20]$ for $\mdm \in$ [$1-10^3$] MeV 
($T_{fo}$ being the freeze-out temperature).

\vs
We will use the previous equations only for particles heavier than 
$\mdm \gtrsim$ MeV (so that they stop annihilating before primordial nucleosynthesis).  
The mixed damping regime is expected to take place for DM particles with an elastic scattering 
cross section 
\beqn
\cs  &\gtrsim& 2 \ 10^{-44} \left(\frac{\mdm}{\U{MeV}{}}\right) \U{cm}{2}, \nonumber \\
\mbox{or} \nonumber \\ 
\bel &\gtrsim&  10^{-56} \left(\frac{\mdm}{\U{MeV}{}}\right) \U{cm}{5} \U{s}{-1} \nonumber 
\eqn 
\noindent 
(bearing in mind that $\cs$ cannot be too large so as to satisfy eq.\ref{up}).

\vs
The elastic scattering 
cross section of heavy DM particles (with $\mdm \gg m_e$)  with neutrinos through 
the exchange of heavy particles $X$ (with a mass $m_X \gtrsim m_W$) is however likely to be  
such that $\sigma_{dm-\nu} \ll \sigma_{\nu-e}$. Since this is in contradiction with eq.\ref{down}, 
one expects the mixed damping effect 
to be  of interest only in non standard circumstances like resonant effects, light DM particles 
etc.

\vs
\subsection{Expressions of the damping lengths}
We can now express the different damping lengths in terms of the DM cross sections. 
Note that we assume $n=2$, $a_{nr} \leq a_{dec(dm-\nu)} \leq a_{eq}$, 
$\tilde{\Gamma}_{\nu-dm} < \tilde{\Gamma}_{\nu-e}$ and we make the assumption that 
DM annihilates. 

\beqn
l_{cd(dm-\nu)} &\sim&  l_{\nu-e} \, \left(\frac{\adec}{a_{dec(\nu-e)}}\right)  \left(\frac{\tilde{\Gamma}_{dm-\nu}}{\tilde{\Gamma}_{(\nu-e)}}\right)^{1/2}_{t_{dec(dm-\nu)}} \nonumber\\ 
&\thickapprox&  l_{\nu-e} \ \left(\frac{\adec}{a_{dec(\nu-e)}}\right)^{5/2}_{ \ a_{dec(dm-\nu)} < a_{dec(\nu-e)}} 
\nonumber\\ 
&\lesssim&  \ 100 \U{pc}{} \ {( 10^{53} \ b_{el})}^{5/8} \left(\frac{\mdm}{\U{MeV}{}}\right)^{-5/8} \nonumber
\eqn
\beqn
l_{md} &\sim& l_{\nu-e}  \ \left(\frac{\adec}{a_{dec(\nu-e)}}\right) \mbox{if} \ a_{dec(dm-\nu)} \gtrsim a_{dec(\nu-e)}
\nonumber\\ 
&\sim& 100 \ \U{pc}{} \ (1.4 \ 10^{54} \, b_{el})^{1/4} \,  \left(\frac{\mdm}{\U{MeV}{}}\right)^{-1/4} \nonumber\\ 
&&\nonumber\\ 
l_{fs(dm)} &\thickapprox& 200 \, \U{pc}{} \, \left(\frac{\mdm}{\U{MeV}{}}\right)^{-1/2} \, 
\left(\frac{a_{dec(dm)}}{a_{dec(\nu-e)}}\right)^{1/2} \ \nonumber\\ 
&&\nonumber\\ 
l_{sd(dm)} &\lesssim& 200 \, \U{pc}{} \, \left(\frac{\mdm}{\U{MeV}{}}\right)^{-1/2} \, 
\left(\frac{a_{dec(dm)}}{a_{dec(\nu-e)}}\right)^{1/2} \ 
\nonumber
\eqn

\vs
As a comparison, in a non-standard scenario where $T_{dec(\nu)} \equiv T_{dec(\nu-dm)} \lesssim 1$ MeV 
and  $\Tdec < T_{dec(\nu-dm)} $, the mixed damping would start at $T_{dec(\nu-dm)}$ instead of $T_{dec(\nu-e)}$. 
This is important as most of the values of $\cs$ one could be tempted 
to consider may yield the scenario $T_{dec(\nu-dm)} < T_{dec(\nu-e)}$. Indeed,  
only a small range of values of $\cs$ can satisfy the conditions eq.\ref{cmd1} and 
eq.\ref{cmd2} associated with the mixed damping regime due to DM collisions with ``standard'' neutrinos. 
Note also that the case $T_{dec(\nu-dm)} < T_{dec(\nu-e)}$ might have a significant 
impact on nucleosynthesis, so one has to treat it carefully. 

\section{WIMP damping mass}

To get quantitative estimates of the damping lengths, we need to specify the DM characteristics. 
In particular, we need to determine when DM decouples 
from neutrinos and electrons (\ie $t_{dec(dm-\nu)}$ and $t_{dec(dm-e)}$). 

\vs 
Similarly to 
$\Tdec$, the decoupling temperature of DM from electrons, if $\mdm > T_{dec(dm-e)} > m_e $, is given by: 
$T_{dec(dm-e)} = \left(\frac{T_0 \, \tilde{H}_r \, \mdm}{ b_{el(dm-e)}  \, \tilde{n}_{e}}\right)^{1/(n+2)} \U{cm}{-1}$ 
(where $\sigma_{dm-e}$ is expected to be temperature dependent with $n=2$). 
However, $T_{dec(dm-e)} < m_e $ is unlikely as this  
would require very large values of $\sigma_{dm-e}$ to compensate the fact that 
electrons become non relativistic and drastically annihilate below their mass threshold. 

\vs
\vs
More details are given in the next section but, generically,  
the elastic scattering cross section of fermionic or bosonic DM candidates with neutrinos or electrons 
can be written as: 
\beqn
\sigma_{dm-\nu} &\propto& \frac{ \ T^2}{(\mdm^2 - m_{S \nu}^2)^2} \nonumber \\
\sigma_{dm-e} &\propto& \frac{  \ T^2}{ (\mdm^2 - m_{S e}^2)^2}\nonumber 
\eqn 
\noindent where the DM couplings to neutrinos or electrons are ``hidden'' in the proportional factor 
and with $m_{S \nu}$, $m_{S e}$ the masses of the exchanged particles 
(they are  bosonic  if DM is a fermion and fermionic if DM is a scalar). 
Note that 
$m_{S e}$ (if charged) is necessarily larger than a few 100 GeV as no charged particle 
has been detected in past accelerator experiments, while 
$m_{S \nu}$ can be below $100$ GeV, 
provided one makes sure that these extra neutral particles are consistent with
existing limits (they should be for example heavier than $\sim \, 45$ GeV 
if significantly coupled to the $Z$ to avoid an anomaly in the $Z$ decay width).

\vs
When $m_{S \nu}$ and $m_{S e} > \mdm$, we find typically for a fermionic or bosonic DM: 
\beqn
\bel \propto  10^{-54} \, \, \left(\frac{m_{S \nu}}{100 \, \U{GeV}{}}\right)^{-4} \U{cm}{5} \U{s}{-1}  \nonumber \\
b_{el(dm-e)} \propto  10^{-54} \,  \,  \, \left(\frac{m_{S e}}{100 \, \U{GeV}{}}\right)^{-4} \U{cm}{3} \U{s}{-1}, \nonumber
\eqn
leading to 
$$
 T_{dec(dm-\nu)} \gtrsim  \U{MeV}{} \left(\frac{m_{S \nu}}{100 \, \U{GeV}{}}\right) 
\left(\frac{\mdm}{\U{MeV}{}}\right)^{1/4}.$$
\noindent 
$T_{dec(dm-\nu)}$ and $T_{dec(dm-e)}$ should actually differ by a factor proportional to 
a certain power of the couplings but we do not expect $T_{dec(dm-e)}$ to be $\ll 1 \U{MeV}{}$.  
Also these temperatures are expected to be slightly different depending on whether DM is bosonic 
or fermionic. 

\vs
There is an exception to these formulae, however, in the case of light DM particles (with a mass below 100 MeV) which 
are required to get an annihilation cross-section dominated by a term in $v^2$ to satisfy 
gamma ray fluxes \citep{bens}; there may be other ways out but this is the simplest one. 
This implies, as explained in the subsection 3.2, that $b_{el(dm-\nu,e)}$ depends rather on $m_{dm}$ 
instead of $m_{S \nu,e}$.

\vs
In case of a mass 
degeneracy $m_{S \nu}$ or $m_{S e} \sim \mdm$, these decoupling temperatures become 
smaller, potentially making the mixed damping relevant. However, even if there exists a 
mass degeneracy large enough so that one would \textit{a priori} expect 
$T_{dec(dm-e)}$ to fall below 1 MeV, DM is likely to decouple from electrons at $T \gtrsim m_e$ 
(due to the drastic change in the electron number density after $e^- e^+$ 
annihilations).

\vs
The fact that $b_{el(dm-e)}$ and $\bel$ appear very close indicates that the DM thermal decoupling may 
be given by  $dm-\nu$ interactions instead of $dm-e$. It would be wrong, or at least dangerous, 
to focus only on dm-e interactions as the $dm-\nu$ interactions determine whether the mixed damping regime 
exists or not.

\vs
Before computing the damping lengths of a typical fermionic candidate (say neutralinos) and a scalar candidate, 
we just mention that, in a situation where there is no mixed damping, 
the largest damping appears to be the free-streaming. The latter is  indeed expected to erase 
all primordial fluctuations with a mass below:  
$$ M_{fs} \propto \ M_{\odot} \ c \ \left(\frac{\mdm}{\U{MeV}{}} \right)^{-15/8} 
\left(\frac{m_S}{100 \U{GeV}{}} \right)^{-3/2}$$
\noindent where $c$ denotes the couplings (at a certain power).

\subsection{Neutralinos}
The expected mass for neutralino DM 
lies between [O(GeV),O(TeV)] \citep{coab}. The lower limit set by the LEP experiment however is 
$\sim$ 37 GeV \citep{alephchi} but this has been obtained 
by assuming a gaugino mass  unification. Also within the 
Minimal Supersymmetric Standard Model (MSSM), the upper limit set by relic density calculations 
based on stau co-annihilations is about $O(400)$ GeV \citep{coae}.  Thus, 
one may find interesting to look at the damping mass of primordial fluctuations 
associated with neutralinos in the mass range [$O(40)$ GeV, $O(400)$ GeV]. (Looking at lighter neutralinos would actually 
be very interesting too, but if they are 
lighter than 10 GeV, one needs to make sure that their annihilation cross-section is suppressed 
by the square of the DM velocity to satisfy radio fluxes \cite{bens}, 
a condition which can perhaps be achieved by introducing, for example,   
a new light gauge boson that is not included in the Standard Supersymmetric Model.) 

\vs
Let us therefore focus on neutralinos heavier than 40 GeV. 
The damping induced by neutrinos is basically given by the interactions 
$\chi_0 \ \nu \rightarrow \chi_0 \ \nu$ through the exchange of sneutrinos. 
The latter has been searched in LEP experiments, notably from single photon 
events, and a limit of $\sim$ 84 GeV has now been set on its mass (under 
the assumptions of gaugino and sfermion mass unification as well as 
no sfermion mixing \citep{alephsnu}). 

\vs
The total elastic scattering cross section is given by: 
\begin{equation} 
\sigma_{\chi-\nu} \thickapprox \frac{ c_l^4 \ T^2}{16 \pi \ (m_{\chi}^2 - m_{\tilde{\nu}}^2)^2}, \label{chinu}
\end{equation}
with $c_l$ the coupling between sneutrino, neutrino and neutralino.
There are two cases for which 
it is interesting to compute the damping scale, namely:
\begin{itemize}
\item a heavy or light neutralino much lighter than sneutrinos, ($m_{\tilde{\nu}} > m_{\chi}$),
\item a heavy or light neutralino degenerate  with sneutrinos ($m_{\tilde{\nu}} \simeq m_{\chi}$) 
so as to benefit from  a resonance that would enhance the cross section and therefore the damping effect.  
\end{itemize}

\vs
Let us first consider a case where there is no mass degeneracy. One gets 
$$\bel \thickapprox c_l^4 \ 10^{-54} \ \left(\frac{m_{\tilde{\nu}}}{100 \ \U{GeV}{}}\right)^{-4} \ \U{cm}{5} \ \U{s}{-1},$$ 
which appears to be much smaller than the value required to have mixed damping. 
So free-streaming is the most relevant effect, erasing up to:
$$ M_{fs} \thickapprox  \ c_l^{3/2} \ \ M_{\odot} \ \left(\frac{\mdm}{\U{MeV}{}}\right)^{-15/8} \ 
\left(\frac{m_{\tilde{\nu}}}{100 \ \U{GeV}{}}\right)^{-3/2} $$
\noindent \textit{i.e.}  $$M_{fs} \thickapprox 8 \ 10^{-10}-10^{-11} \ \left(\frac{c_l}{0.3}\right)^{3/2} \ 
\left(\frac{m_{\tilde{\nu}}}{100 \ \U{GeV}{}}\right)^{-3/2} \ M_{\odot}$$ for 
$m_{\tilde{\nu}} > m_{\chi}$ and $m_{\chi} = 40- 400$  GeV.  
This is to be compared with $$M_{cd} \thickapprox 5 \, 10^{-18}- 4 \ 10^{-16} \ M_{\odot} \ 
\left(\frac{c_l}{0.3}\right)^{15/2} \left(\frac{m_{\tilde{\nu}}}{100 \ \U{GeV}{}}\right)^{-15/2}$$ 
for $m_{\chi} = 40-400$ GeV. All these values are actually far from the calculation done by \citep{bst}.

\vs
In case of a mass degeneracy, on the other hand,  
the $\chi_0-\nu$ elastic cross section, 
$$\sigma_{\chi-\nu}  \propto  \frac{c_l^4 \ T^2}{\left(-m_{\tilde{\nu}}^2 + m_{\chi}^2 + 2 T m_{\chi}\right)^2} \ , $$
can become temperature independent provided $m_{\chi}^2 -m_{\tilde{\nu}}^2  < 2 \, T \, m_{\chi}$.   
Let us define $m_{\tilde{\nu}}$ as $ m_{\tilde{\nu}} = m_{\chi} (1 +\delta)$. So  
\begin{itemize}
\item for $T > (2 \, \delta + \delta^2) \, m_{\chi}/2$, the cross section is temperature independent: 
\begin{equation}
\sigma_{\chi-\nu} \propto \frac{c_l^4 }{4 \ m_{\chi}^2}, \ 
\label{cs1}
\end{equation}
\item while for $T < (2 \, \delta + \delta^2) \, m_{\chi}/2$, the cross section is temperature dependent:
\begin{equation}
\sigma_{\chi-\nu} \propto  \frac{c_l^4  \ T^2}{m_{\tilde{\nu}}^4}. \  \ 
\label{cs2}
\end{equation}
\end{itemize}
\noindent For example,  one expects $\sigma_{\chi-\nu}$ to be constant at  temperatures 
 above 10.5 GeV, if one assumes $m_{\chi} \sim 100$ GeV and 
$\delta=0.1$. 

\vs
Thus, unless the degeneracy appears to be adequate so that DM can decouple at $T \lesssim 1$ MeV, 
the damping generated by the 
dm-$\nu$ collisions should eventually 
be computed in two steps. First using formula \ref{cs1} up to $T_c = (2 \, \delta + \delta^2) \, m_{\chi}/2$ 
and then using eq.\ref{cs2} up to $\Tdec$ (where $\Tdec$ will be estimated by use of 
eq.\ref{cs2}). One nevertheless expects the damping to be dominated by the late times 
so using eq.\ref{cs2} only should finally give a good estimate.   

\vs 
One can compute the neutralino thermal decoupling temperature in ordinary situations 
from eq.\ref{chinu}, using eq.\ref{calc1} and eq.\ref{calc3}.  
One finds 
$$ \Tdec \sim 5.3 \ \U{MeV}{} \ \left(\frac{c_l}{0.3}\right)^{-1} \,  
\left(\frac{m_{\tilde{\nu}}}{100 \U{GeV}{}}\right) \, \left(\frac{\mdm}{\U{MeV}{}}\right)^{1/4},$$ 
for $T < (2 \, \delta + \delta^2) \, m_{\chi}/2$ 
(\ie 75 MeV for $\mdm =$ 40 GeV and 133 MeV for $\mdm =$ 400 GeV). 
A similar temperature is expected for $T_{dec(dm-e)}$ as the electrons are 
relativistic above $m_e$ and behave like neutrinos. 
 On the other hand, if the degeneracy was about $10^{-3}$ for $\mdm =100$ GeV (\ie $\delta =10^{-5}$, 
$m_{\tilde{\nu}} = 100.001$), then only one step would be necessary to compute the thermal 
decoupling epoch as the cross section remains constant up to $T \lesssim $ MeV. 
As a result, $\Tdec$ could be less than $\sim$ 1 MeV for 
$m_{\chi} \sim$ 100 GeV, which would further considerably increase the damping mass (at least up to 
$ 1 \ M_{\odot}$), although such a degeneracy cannot be taken seriously.

\ghost{
\begin{equation}
\begin{array}{l|l|l|l|l|l}
\hline
 &M_{cd(\nu)} &M_{md} &M_{fs} &M_{sd} &M_{ann} \\
\hline
\mdm < m_F &&&&& \\
\hline
\mdm \sim m_F &&&&& \\
\hline
\end{array}
\end{equation}}

\subsection{Light DM}

Although the most popular DM candidate certainly appears to be a ``heavy'' neutralino,  
the possibility of having light DM candidates may still be a viable solution.
In particular, if not coupled to the $Z$, light scalar particles (with $O(\U{MeV}{}) \lesssim \mdm \lesssim O(10 \U{GeV}{})$)  
may be solution to the DM issue, provided they are coupled to a light gauge boson $U$ (with a 
mass $m_U \gtrsim \mdm$) or to heavy fermionic particles $F$ (although in this case, the gamma ray, radio flux 
as well as $g-2$ constraints might 
impose that DM be made of non self-conjugate particles and require 
to introduce a set of new 
particles in order to kill the too large $F$ contribution to the muon and electron $g-2$). 
  
\vs
When coupled to fermionic particles $F$, light scalars are not expected to experience a large 
damping effect because of the  mass difference between DM and the $F$ particles  
(supposed to be very heavy from past accelerator experiments).   
There is an exception, however, if DM mainly annihilates  through a very light 
particle $F^0$ (with a mass $m_{F^0} \sim \mdm$) which is not significantly coupled to the $Z$ 
(and which would have escaped accelerator limits).

\vs
On the other hand, if DM is coupled to a new and light gauge boson $U$, one expects 
damping effects to be significantly  enhanced. 
The elastic scattering cross section 
$\sigma_{dm-\nu}$ (associated with the exchanged of a $U$ boson through a t-channel) is given by: 
$$ \bel = 10^{-33} C_U^2 \ f_{U_l}^2 \ \left( \frac{m_U}{\U{MeV}{}}\right)^{-4} \U{cm}{5} \U{s}{-1}.$$ 
In \cite{bf}, it is shown that -- to satisfy both relic density and $g-2$ constraints  -- the product of 
the couplings $C_U$ and $f_{U_l}$ (which correspond to 
the $U$ boson couplings to DM and to ordinary particles respectively) 
must satisfy the relationship 
$C_U \ f_{U_l} = (3-12) \ 10^{-8} (\frac{\mdm}{\U{MeV}{}})^{-1} \ (\frac{m_U}{\U{MeV}{}})^2$ 
so $$ \bel \sim [9-144] \ 10^{-49} \ \left(\frac{\mdm}{\U{MeV}{}}\right)^{-2}\U{cm}{5} 
\U{s}{-1}$$ (where we assume ``universal'' $f_{U_l}$, \ie same value for $U-e-\bar{e}$ 
and $U-\mu-\bar{\mu}$ for instance).
This implies a decoupling scale-factor: 
$$ \adec \sim (5-10)  \ 10^{-9} \  \left(\frac{\mdm}{\U{MeV}{}}\right)^{-3/4}$$
which is independent of $m_U$ and maximal for small values of the DM mass. 
As an example, the decoupling epoch for $\mdm = 10$ MeV 
corresponds to $\adec \, \sim  (0.8-1.5) \ 10^{-9}$ 
which implies a mixed damping mass of about 
$$M_{md} \sim 21-172 \ M_{\odot} \ (\mbox{for} \ 10 \U{MeV}{}).$$ 
This damping mass obviously gets smaller for larger DM mass and it is worth to mention that 
 the cross section for $\mdm=10$ MeV is close to the upper limit that arises from the condition 
of free-streaming neutrinos, namely 
$\bel \lesssim 1.86 \, 10^{-56} \, \left(\frac{g_e}{g_{dm}}\right)^2 \, \left(\frac{\mdm}{\U{MeV}{}}\right)^{-1} \, 
x_d^{-3} \, e^{2 x_d} \ \U{cm}{5} \U{s}{-1}$ (so $\mdm =10$ MeV is close to the upper limit 
for $ \bel \sim 144 \ 10^{-49} \ \left(\frac{\mdm}{\U{MeV}{}}\right)^{-2}\U{cm}{5} \U{s}{-1}$).
Note that one would naively expect a bigger effect in the case of light DM due to the smallness of $\mdm$ 
but the latter is actually  compensated by the smallness of the product $C_U \ f_{U_l}$. 

\vs
With the expression  $\adec$ obtained above (valid only if DM particles are coupled to a $U$ boson), 
one can readily see that the mixed damping regime is at work only for a DM mass in the  range 
$\mdm \in [\sim 5 \, , (500-700)] \, \U{MeV}{}$ (heavier DM would decouple before 1 MeV, lighter particles 
would give rise to the alternative scenario where the neutrino decoupling occurs at $T_{dec(\nu)}<$ MeV). 

\vs
In the previous examples, however, we assumed that the number density of DM particles 
was equal to the number density of anti DM particles. If one relaxes this assumption (assuming non self-conjugate DM), 
then $C_U$ can take much larger values (but likely to be $\in [10^{-3},O(1)]$). As a result, the 
decoupling scale-factor which is proportional to $\bel^{1/4} \propto (C_U  \ f_{U_l})^{1/2}$ 
can become  at most$\sim \sqrt{10^3}$ larger than our previous estimate, so that the damping mass 
that one expects in such situations could possibly reach $\sim (\sqrt{10^3})^3 \sim 3 \ 10^4$ times the 
mass $M_{md}$ estimated without the  number density asymmetry. But a change in $C_U$ is, in fact, 
particularly relevant 
for DM particles with $\mdm \gtrsim O(10)$ MeV, for example, as  
the upper limit on the cross section (set by the condition of free-streaming neutrinos) 
may remain possible to satisfy. 

\vs
Such particles, however, would not have any 
residual annihilations so they would not produce gamma rays but on the other hand, 
they would have an oscillating linear matter power spectrum largely damped below 
$O(10^3 M_{\odot})$, which finally provides a  signature.

\vs
One can also consider larger values of $f_{U_l}$ by 
relaxing the assumption of universal couplings. Indeed, the values mentioned before for  $C_U  \ f_{U_l}$ 
actually supposed that the coupling $dm-\nu-\bar{\nu}$ is of the same order of magnitude as  
the couplings $dm-\mu-\bar{\mu}$ or $dm-e-\bar{e}$ (constrained through the $U$ contribution to 
the muon and electron anomalous magnetic moments). Writing $f_{U_l}^{\nu} = x 
f_{U_l}^{e}$ (with $x \geq 1$)
($f_{U_l}^{e}$ being the relevant coupling for low DM mass), 
one finds that the damping mass is increased by a factor $x^{3/2}$ and $a(t)$  
by a factor $x^{1/2}$ (still provided one makes sure that 
the condition of free-streaming neutrinos is satisfied).

\vs
Such situations may finally allow for DM particles heavier than the limit $500-700$ MeV to experience 
a non-negligible mixed damping (while they were not supposed to with smaller values of the couplings). 
Indeed, an increase in the couplings simultaneously 
allows for eq.\ref{down} to be satisfied and for $M_{md}$ to be enhanced.

\vs
Note that the mixed damping mass is not expected to become larger in case of a degeneracy between $m_U$ and $\mdm$, 
because the interaction at the origin of the damping effect proceeds through a t-channel. 
On the other hand, such a degeneracy would affect the decay modes of the $U$ boson which would then 
decay into ordinary particles only (\ie into $e^+ e^-$ and $\nu \bar{\nu}$ instead of dm dm$^\star$, 
although one has to make sure this is not excluded 
by nuclear experiments).

\vs
The previous examples provide a theoretical framework to investigate the 
effect of a ``collisional'' cut-off in the linear matter power spectrum at scales 
$\sim \ 10^3 \ M_{\odot}$. This is actually too small compare to our resolution 
 but the question of whether a cut-off in the linear matter power spectrum 
is still present in the $P(k)_{nl}$ can be partially answered by investigating 
the effect of a cut-off at $10^6-10^9 \ M_{\odot}$. 

\vs
The main feature, say oscillations, in the light DM linear matter power spectra should be 
somewhat similar to that obtained in \cite{brhs} (although the damping mechanisms are different).  
However, to extend our point to other class of DM candidates, we decide to perform simulations 
of collisionless WDM and see whether a cut-off in the linear matter power spectrum at large 
scale ($10^6-10^9 \, M_{\odot}$) also appear in the $P(k)_{nl}$ and at which redshift. If such a cut-off 
appears difficult to detect, then non conventional candidates as proposed in this paper 
will be very difficult to exclude from the measurement of the non linear matter power spectrum 
only.

\section{Regenerating small scale power \label{hugues}} 

Estimations of the matter power spectrum 
on scales $l \gsim 200$ \hkpc are now routinely derived from 
weak and strong lensing techniques at  low redshift, 
or indirectly from the correlation of spectra of the Ly-$\alpha$ forest at $z\sim3$. 
Fair agreement of the observations with the slope of the 
non-linear power spectrum expected in the CDM paradigm 
is commonly regarded as a confirmation of the assumed 
``cold'' nature of dark matter. This is a biased conclusion, as the 
 $n_{\rmn{nl}}=-1.4$ slope of $P(k)_{\rmn{nl}}$ at these small scales 
($n=\rmn{d}\,\ln{P}/\rmn{d}\,\ln{k}$) and at this low redshift 
range is known to behave as an ``attractor'' to the evolution 
of $n$, for a variety of DM models \citep{Sco96}.

Using numerical simulations of structure formation 
with initial power spectrum restricted to two separated 
scale ranges, \citet{Bag97} demonstrate that  
 the non-linear growth of the small scale modes is mostly driven by 
the cascade of power due to the coupling with larger scale modes. 
As a result, DM models with very different $P(k)_{\rmn{l}}$ on small scales but similar large scale 
power are expected to yield small-scale $P(k)_{\rmn{nl}}$ of the same shape.  
Focusing on measurements of the Ly-$\alpha$ forest, \citet{Zal03} 
make the same point at a somewhat more observational degree, and 
clearly show how  little information is obtained on the initial shape of $P(k)_{\rmn{l}}$ 
at $k\gsim 10$ \hMpcInv by inverting the flux power spectrum, as long as one 
starts with a sufficiently large prior in $n_{\rmn{l}}$.

\vs

Physically, the IDM and WDM models considered here 
are as well motivated as CDM (but for a little more complexity), 
and they are \emph{a priori} as probable as the latter.  It is 
therefore necessary to verify if, when and at which scale their specific 
matter power spectrum becomes sufficiently close to that of 
CDM so that current observations are unable to separate between the two.

\vs

To assess this, we simulate the gravitational formation of structure using $128^3$ 
collisionless particles in a small, 5  \hMpc side comoving box, 
with the cosmological parameters measured by \emph{WMAP} \citep{wmap}, 
starting from z=100. The mean inter-particle distance is 39 \hkpc and 
our force resolution is one tenth this value. We evolve the particle distribution 
with the public version of {\sc gadget} \citep{SprGadget2001}. 
We consider a CDM and four WDM models with warmon masses 0.6, 1.1, 2 
and 3.5 keV: the associated free-streaming lengths correspond 
to Lagrangian masses $10^{9}$, $10^{8}$, $10^{7}$ and $10^{6}$ \hmsun respectively. 
The CDM and 0.6 keV WDM models bracket the series of models discussed in this work (including IDM), 
while the other 3 samples are intermediate cases.  We keep 
the same phases for all the modes in the five simulations. In 
 the following discussion we will not tackle the issue of cosmic variance.  
We normalize the modes $\delta_{k}$ inside the box so 
that they correspond to $\sigma_{8}=0.9$.  In doing so, 
we neglect the effect of primordial fluctuations 
with comoving scales greater than the box size: the power effectively realized 
in the box is 3 percent of the total cosmological power.

\vs

To estimate the dispersion expected if we were to include  
all the modes up to the scale of cosmic homogeneity, 
we have performed two series of similar simulations which we 
call test simulations. Checking this effect is  
necessary because of the combination of our 
limited dynamic range in scale and of our choice of a small simulation box.

\vs

The first series of test simulations consists of a CDM and a 0.6 keV WDM model with 
the same phases as above in a flat cosmology with the same and $\sigma_{8}$ and $t_{0}$ but with 
different $\Omega_{0}$: $\Omega_{0}$=0.15 and 0.45, corresponding 
to a $\delta_{> 5\hMpc}=\pm\sigma_{5}/3$ background and $\Omega_{0}$=0.03 and 0.57, 
corresponding to a $\delta_{> 5\hMpc}=\pm\sigma_{5}$ background. In these simulations, 
$h$ varies accordingly to ensure the same age $t_{0}$ as the cosmological 
model preferred by the \emph{WMAP} results. Recall that we take $\sigma_{8}=0.9$ and we 
define $\sigma_{5}$ as the RMS density fluctuations in spheres of radius 5 \hMpcDot 
This series of simulation checks the effect of large scale bias in a peak-background split approach. 

While we find  the regeneration of small scale power for our 0.6 keV WDM model 
to be very similar in the $\Omega_{0}$=0.45 and $\Omega_{0}$=0.57 simulations  
to the result of the reference mean density simulation, the low density simulations 
(especially $\Omega_{0}$=0.07) show that small scale regeneration occurs 
significantly earlier when one evolves in an underdense, 5 \hMpc region of the universe 
compared to a mean or high density region. This is expected, because for the same absolute 
normalisation of the power spectrum, the growth of modes occurs earlier in low density flat cosmologies. 
As a result, the cascade of power from the larger modes that we describe in detail below 
also occur earlier in low density cosmologies and the 0.6 keV WDM power spectrum 
catches up earlier on with the CDM power spectrum. However, this phenomenon 
will only strengthen our conclusions.

The second series is the same CDM and 0.6 keV WDM model as in our main series, 
changing only the box size from 5 to 20 \hMpc (and the mean 
inter-particle separation to 156 \hkpcKD This second series 
verifies that enough large-scale power is already realized inside the 5 \hMpc box 
for the same cascade of power to occur as would happen in a larger box.  
We have found that  the neglected larger scale modes have little impact on our conclusions: 
the dispersion due to fluctuations in matter density is smaller than the differences we discuss in the power 
spectra of different DM models, and the regeneration of small-scale power is the same 
in the 5 and 20 \hMpc boxes at the overlapping  wavelengths. 
For completeness sake we point out that we did find reduced small-scale regeneration in a 1 \hMpc box simulation because of 
 lack of large scale power in this case. This allows us to conclude that  
 5 \hMpc is larger than, but close to the minimum box size necessary for 
 our conclusions to be meaningful. 

\vs

\begin{figure*}
  \begin{center}
    \includegraphics[width=19cm]{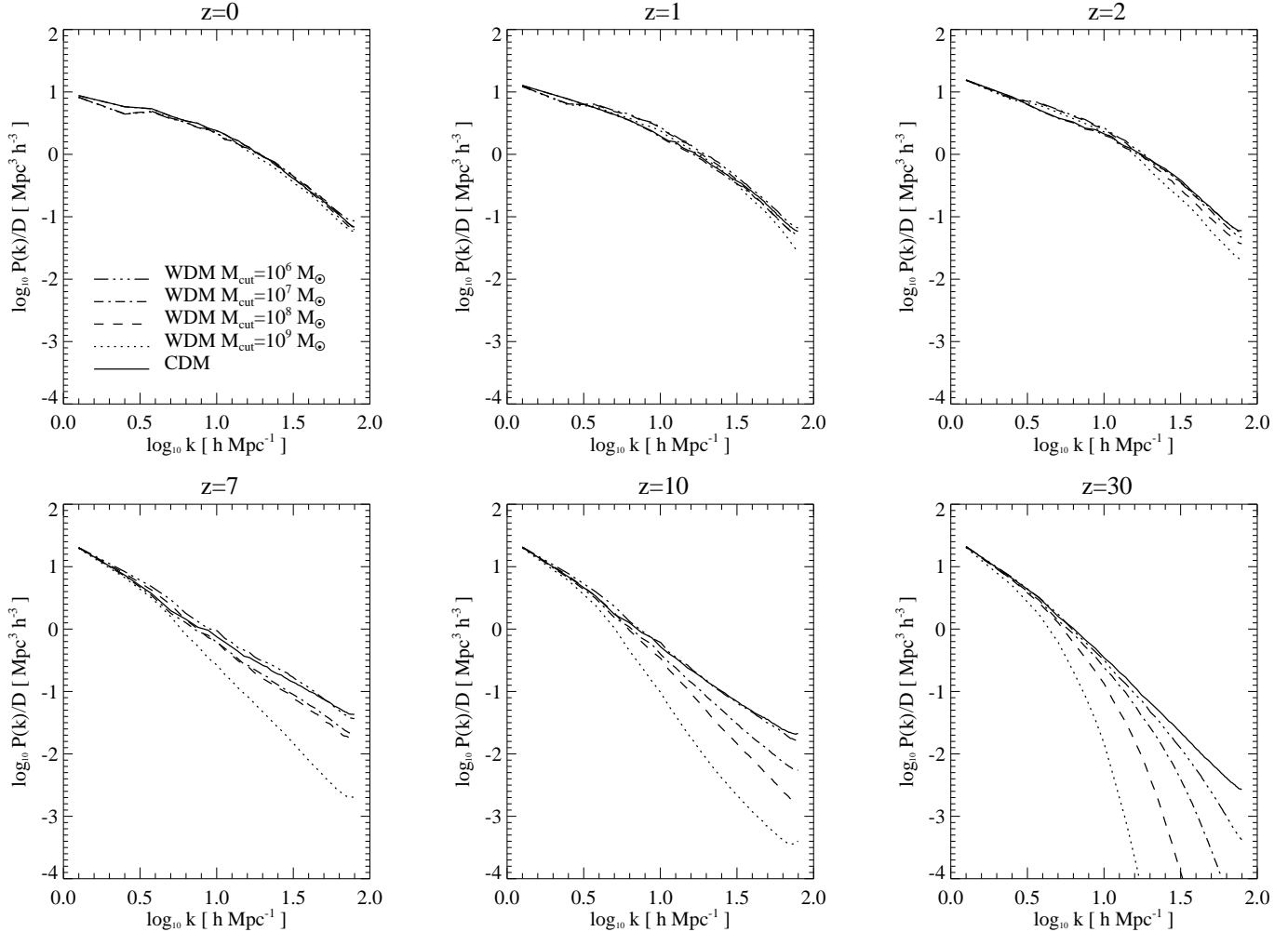}
  \end{center}
  \caption{\label{fig:Power} Snapshots of the dark matter power spectra measured in simulations of a series of DM models.  CDM and 0.6, 1.1, 2 
and 3.5 keV heavy warmon WDM corresponding to respectively $10^{9}$, $10^{8}$, $10^{7}$ and $10^{6}$ 
\hmsun Lagrangian masses in the free-streaming length are shown. The power spectra have been divided by 
the linear growth factor to facilitate the comparison between redshifts. Note the exponential cut-off at z$=30$ 
in the spectrum of WDM models and the similarity of the spectra of all five models over the simulated scales at z$\le$2.}
\end{figure*}

Figure~\ref{fig:Power} gives the matter power spectra $P(k)_{\rmn{nl}}$ of our five simulations, 
from z=30 to z=0. The spectra are computed up to $k_{\rmn{Ny}}=2\,\pi\, 64 / 5\sim 80$ \hMpcInv 
using a deconvolved TSC scheme. To facilitate the comparison, 
they have been divided by the normalized linear growth factor 
$D_{z=z_{\rmn{init}}\rightarrow z=z_{\rmn{plot}}} / D_{z=z_{\rmn{init}}\rightarrow z=0}$.

The qualitative evolution of the power spectrum is similar to that found by \citet{Kne02} 
in their simulations of WDM models with warmon masses of 0.5 and 1 keV. At z=30, 
shortly after starting the simulations, the shape of the measured spectrum is in all cases close to 
the linear power spectrum of the initial conditions. By z=10, non-linear effects have already 
significantly altered the shape of all power spectra on the smallest scales $k\gsim10$ \hMpcInvDot At z=7, 
WDM power spectra are already reasonably close to CDM, with the exception of the 0.6 keV simulation, 
where the power spectrum is still steeper and keeps the signature of the $10^9$ \hmsun cut-off. At z$<$2 
however, the power spectrum of the 0.6 keV WDM simulation is within a factor of 2.5 of that of the CDM down to 
the smallest scales probed. At z=1, the curves can be considered the same within 10 percent.  
These plots confirm that the late $P(k)_{\rmn{nl}}$ is similar down to scales 
as small as 80 \hkpc at z$\lsim$2 for models with so different input power spectra 
such as CDM and the 0.6 keV WDM scenario. We find the slope of the evolved 
power spectrum to tilt from $n\sim -1$ for $k<10$ \hMpcInv to $n\sim -1.6$ for $k>10$ \hMpcInvDot  
Furthermore, at a fixed comoving scale, the power spectrum of WDM models 
with the larger warmon mass (and smaller free-streaming length) catches up earlier with the 
CDM power spectrum than do models with larger free streaming lengths. The intuitive guess that  
at a fixed WDM model, power on large scales matches earlier CDM power 
than power on a smaller scale is confirmed. This is of course the signature of the exponential cut-off
present in the linear power spectrum.

\begin{figure*}
  \begin{center}
    \includegraphics[width=19cm]{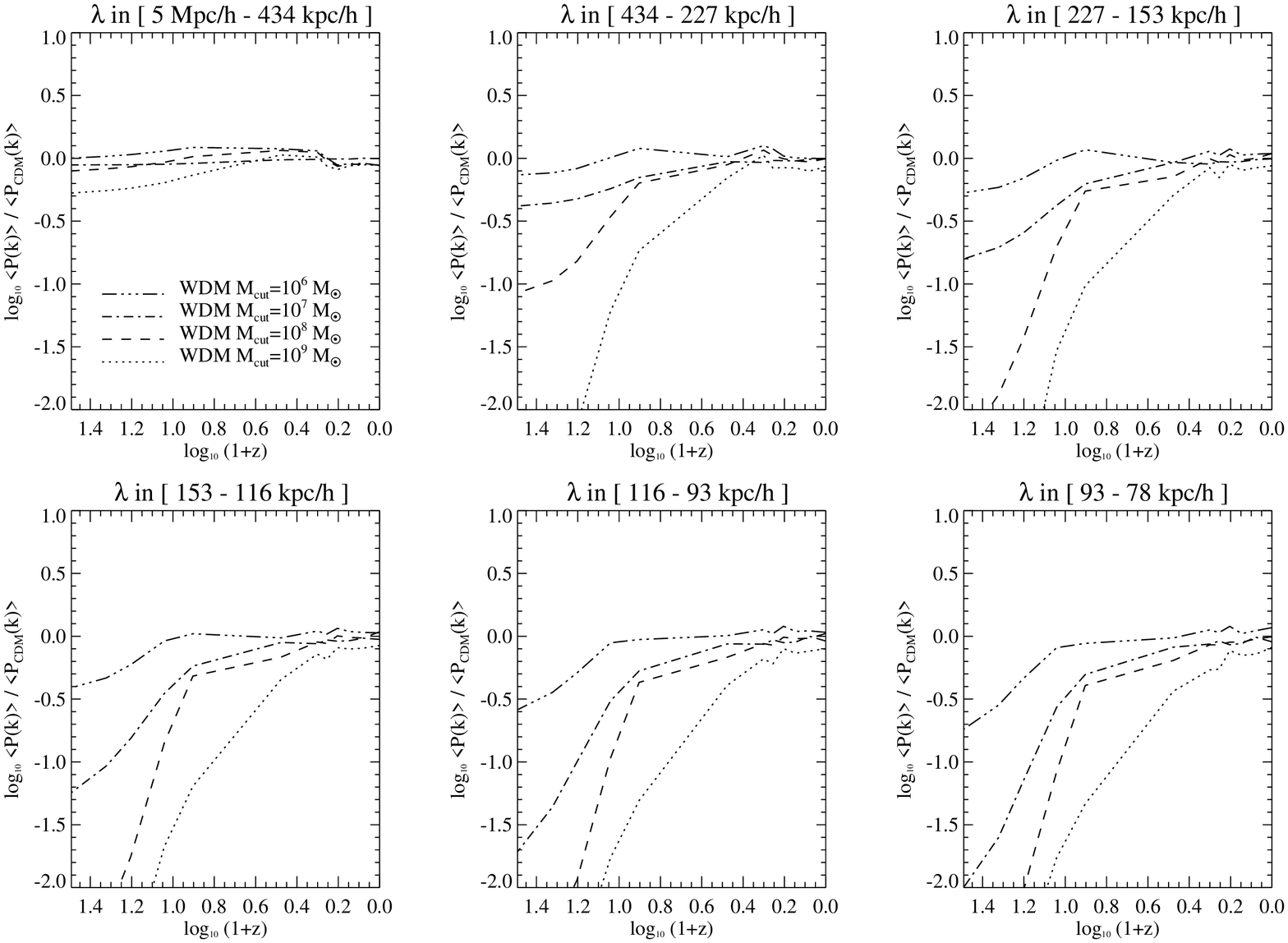}
  \end{center}
  \caption{\label{fig:PowerRatio} Evolution with redshift of the ratio of (1) the mean of the power spectrum 
of each of the four WDM models (0.6, 1.1, 2 and 3.5 keV particles corresponding to respectively $10^{9}$, $10^{8}$, $10^{7}$ and $10^{6}$ 
\hmsun Lagrangian masses) to (2) the mean of the power spectrum of CDM model, in six consecutive  
bins covering the whole wavelength range probed by the simulations. Note how the ratio converges to unity at late 
times for all WDM models in each of the 6 bins. The marginal exception is the evolution of the 0.6 keV WDM model which 
shows 10 percent less average power than CDM on the smallest scales bin at z=0. The power spectrum of this model 
may strongly depart from CDM on scales $\log_{10}(k)\sim2$ even at z=0.}
\end{figure*}

These monotonic behaviours are more clearly seen in Figure~\ref{fig:PowerRatio}. There, we 
first divide our wavelength coverage ($k \in  [0.1 \; 1.9]$ \hMpcInvKet into 6 consecutive bins, 
corresponding to the 6 panels shown. In each bin we estimate the average 
of the power spectrum $\langle P(k)_{\rmn{nl}} \rangle$ for the 
CDM simulation and the four WDM simulations at each redshift. We then divide 
the time evolution of the averaged power spectrum $\langle P(k)_{\rmn{nl}}(z)\rangle$ 
of each WDM simulation by the time evolution of the averaged CDM power spectrum.  
We only discuss the upper left and lower right panels, the other 4 
being intermediate cases shown here for reference. On the upper left panel (bin associated to the largest scales), 
the ratio of the average power measured in the WDM models to the average power in the CDM model is within 10 percent 
of unity from z=30 down to z=0 for all warmons but the lightest, {\em i.e.} the 0.6 keV particle. 
In this last case, the signature of the initial exponential cut-off is already visible on large scales, 
with a factor 2 less power than CDM. This deficit gradually disappears over 10$>$z$>$2. 
The lower right panel (bin associated to the smallest scales that we probe) clearly separates our four WDM models. 
The average power of the 3.5 keV WDM model is the first to catch up with CDM, at z$\sim$12. 
The increase in the ratio of 2 and 1 keV WDM  power to CDM power is abrupt before z$\sim$10, slows down in  
10$>$z$>$2 and becomes comparable to CDM  at low redshift. The ratio of the 0.6 keV WDM to CDM power 
increases less steeply than the 1 and 2 keV power initially, seems to slow down at z$<$1, 
but has not yet exactly matched the CDM power at z=0: there is a remaining 10 percent offset (but probably 
indistinguishable with current observations). 
However, in all these cases, and also including the four remaining panels, 
the ``monotonic'' feature of the process of small scale regeneration is preserved. Figure~\ref{fig:PowerRatio} 
 contains the same information as Figure~\ref{fig:Power}, it only has more sampling points in redshift. We can  
extrapolate that the power of the 0.6 keV model may not have matched the power of the CDM model at z=0, 
if one would look at scales smaller than $\log_{10}{k}=1.9$. As we want to simply illustrate the late-time 
degeneracy in the power spectrum between a 0.6 keV WDM model and a CDM model at scales $\sim 100$ \hkpcCom we leave 
 this issue for further work.

\vs

Regarding the regeneration of small-scale power we find that observations (e.g. lensing) 
probing the matter distribution at z$\lsim$1 at scales as small as 100 \hkpc will be unable to detect 
the signature of the initial exponential cut-off of WDM models with warmon mass as small as 0.6 keV. 
For this warmon mass, the signature of a cut-off will only be clearly visible on lenses at z$\lsim 4$, not even considering 
systematic errors nor cosmic variance. All WDM models with larger warmon mass will be increasingly harder to single out.  
Alternatively, one may need to probe  scales  smaller than 100 \hkpcDot  The converse, and our main point, 
is that finding at the above scales 
a z=0 power spectrum in agreement with that expected from the non-linear 
evolution of an initial CDM model cannot rule out physically motivated 
alternatives such as WDM or a light DM scenario presented in this paper.  

We emphasize that this is true even if we take into account the possible effect of the modes 
with wavelength larger than our simulation box: while evolving in a 1 $\sigma$ 
overdense region (defined with respect to our neglected modes) does not change the regeneration of small scale modes 
compared to what happens in a mean density region, simulating a 1 $\sigma$
underdense region we find small scale regeneration to occur even earlier.

\vs

\begin{figure*}
  \begin{center}
    \includegraphics[width=15cm]{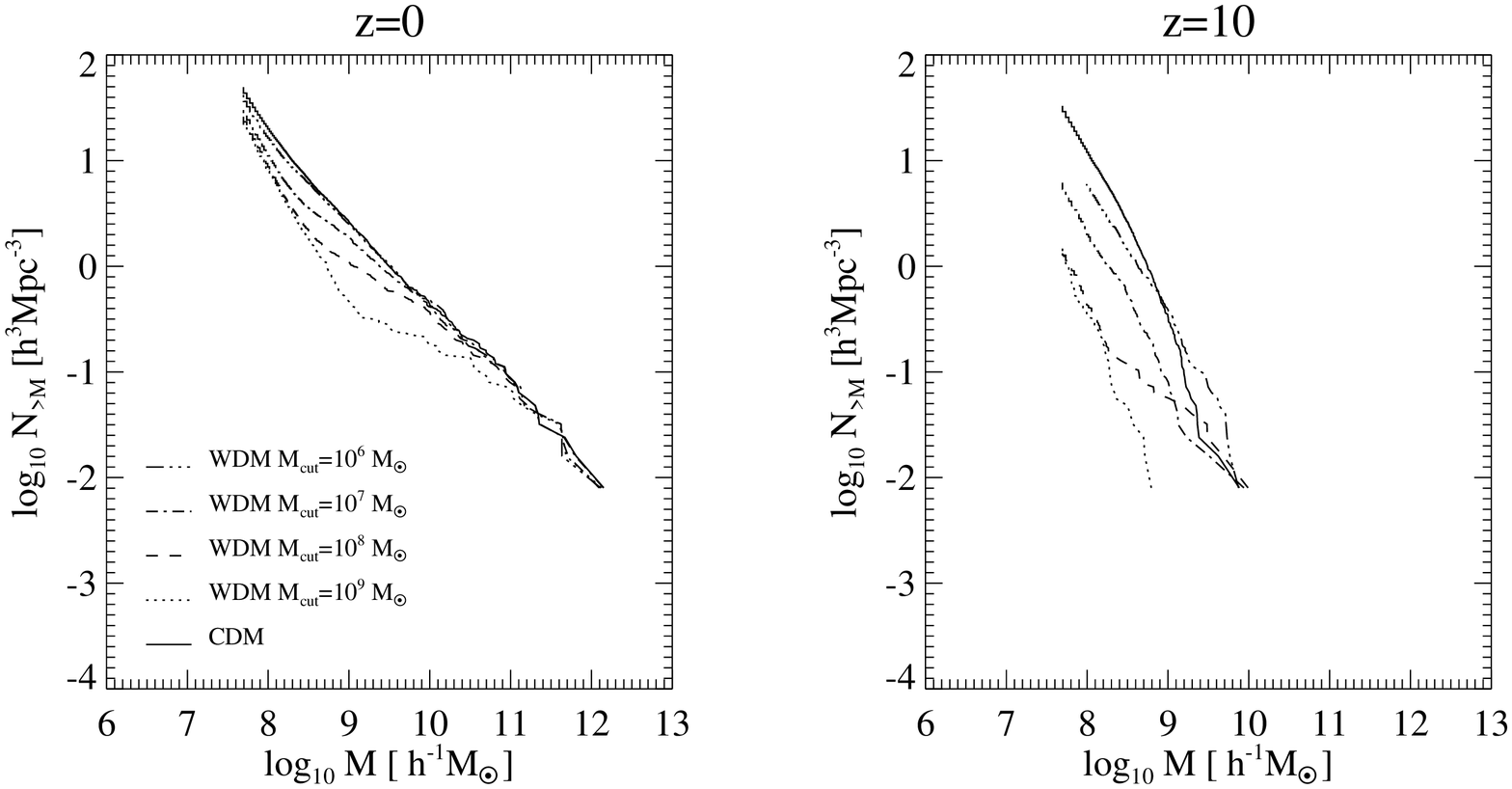}
  \end{center}
  \caption{\label{fig:MF} Early (z=10) and late-time (z=0) dark matter halo mass functions of the CDM 
model compared to those of the WDM models. Note the deficit by a factor of $\sim40$ of 0.6 keV WDM haloes 
at z=10 compared to the abundance of CDM haloes.}
\end{figure*}

To conclude this section, we note that the mass function of collapsed 
objects is also a sensitive probe \emph{at high z} of a possible 
exponential cut-off in the primordial power spectrum. It has the advantage 
over power spectrum measurements that large variations in the abundance of haloes 
at the epoch of the formation of the first stars or later at reionization can be rather 
tightly constrained even with current observations, at a fixed physical prescription 
for the evolution of the baryonic component. 
The drawback resides in first, the current large freedom 
in physical models of reionization or of the formation of the first stars 
 and second, in that the mass function is also very sensitive to a possible 
small-scale primordial non-gaussianity. While we plan to address 
these issues in more detail in the future,  Figure~\ref{fig:MF} gives the z=0 (left panel) and z=10 (right panel) 
mass function of collapsed dark matter haloes in our simulations. Note the mass range probed 
($5\times 10^{5}$ to $10^{12}$ \hmsunKet At z=10, the deficit of haloes of all masses 
is evident in the 0.6 keV WDM model compared to CDM. It is also apparent in the simulated WDM 
models with higher warmon mass, although mainly at the low-mass end of the curves. The z=10  
abundance of the largest, $10^{10}$ \hmsun mass haloes is similar in the CDM and the 1.1, 2, 3.5 keV WDM models. 
At z=0 the abundances of haloes with $M_{\rmn{tot}}\gsim 10^{11} \hmsun$ match in all 5 models. In the range 
$10^{8.5} \lsim M_{\rmn{tot}} \lsim 10^{11} \hmsun$, the mass function of the 0.6, 1.1 and 2 keV models shows 
a deficit of haloes compared to CDM, with the 0.6 keV model showing the lowest abundance. At   
$M_{\rmn{tot}} \lsim 10^{8.5} \hmsun$, halo abundances tend to coincide again.

\vs

\section{Conclusion}
 
In this paper we compare two distinct DM candidates. 
One is the most popular candidate, say the lightest 
neutralino, while the other one is light DM particles.  

\vs
Both these candidates are expected to have weak interactions. They both annihilate 
to get the proper relic density but, in the case of neutralinos, collisional 
damping effects are completely negligible while, in the case of light DM, the collisional 
damping (and more specifically the mixed damping) yields a cut-off in the linear matter 
power spectrum which can realistically reach $10^3 \, M_{\odot}$.   

\vs
Because the existence of this cut-off seems a promising way to discriminate among these two extreme DM scenarios,  
we perform a series of numerical simulations to obtain the corresponding non-linear matter power spectra.    
To make the case even more obvious,  in fact, we simulate scenarios with a more drastic cut-off than what 
is expected in our model. More realistic simulations of light DM scenarios will be published in a forthcoming 
paper but we do expect similar conclusions.      
This enables us to quantify the time evolution of $P(k)_{nl}$ as a function of scale, and  
also to precisely assess to which extent the late time ($z<1$) small-scale  
power spectrum can be used as an efficient tools to distinguish between standard 
(collisionless) CDM and alternative scenarios.  
 
\vs
Since we are limited by the size of our simulations, we 
simulate a cut-off at $10^6,\ 10^7, \ 10^8$ and $10^9$ \hmsunDot
We show that the corresponding $P(k)_{nl}$ is similar to the non-linear matter 
power spectrum of CDM particles for $z< 1$, so weak lensing 
measurements except on the smallest scales (100 kpc/h or less for lenses located at $z<1$) 
are not able to discriminate between usual CDM particles and more exotic candidates. 
Note here that current measurements of the cosmic shear probe the shape of the matter 
power spectrum down to 200 \hkpcCom while planned upcoming surveys 
will go down to 70 \hkpc and start to probe a critical region for a $10^9$ \hmsun cut-off.  

\vs
On the other hand, there exist other sensitive signatures of the nature 
of the dark matter, such as structural parameters, clustering of virialized dark matter haloes and the reionization 
epoch. However these are more indirect constraints and involve poorly understood interactions between baryons and DM, 
which considerably weakens their predictive/discriminative power.

\section*{Acknowledgements}
We would like to thank Brice Menard for useful discussions.
CB is supported by an individual PPARC Fellowship. 
HM acknowledges financial support from PPARC. 
JD is supported by a Leverhulm grant. 


\end{document}